# The role of quantum superposition in a coupled interferometric system for macroscopic quantum feature generations

Byoung S. Ham

Center for Photon Information Processing, School of Electrical Engineering and Computer Science, Gwangju Institute of Science and Technology

123 Chumdangwagi-ro, Buk-gu, Gwangju 61005, South Korea

bham@gist.ac.kr

(Submitted on Feb. 23, 2021)

**Abstract:**
Quantum entanglement is the quintessence of quantum information processing mostly limited to the microscopic regime governed by Heisenberg's uncertainty principle. For practical applications, however, macroscopic entanglement gives great benefits in both photon loss and sensitivity. Recently, a novel method of macroscopic entanglement generation has been proposed and demonstrated in a coupled interferometric system using classical laser light, where superposition between binary bases in each interferometric system plays a key role. Here, the function of path superposition applied to independent bipartite classical systems is analyzed to unveil secrets of quantum features and to convert a classical system into a quantum system without violating quantum mechanics.

**Introduction**

Quantum entanglement [1] has been an essential concept in quantum mechanics over the last century. Recently, the basic concept of quantum entanglement has been implemented for a potential quantum device such as a quantum computer [2-5]. Previously, quantum key distribution had already been applied for secured information communications between two remotely separated parties [6-8]. The third application of quantum entanglement is quantum sensing, where macroscopic quantum entanglement is an essential requirement [9-12]. Although a higher-order entangled photon state, called the N00N state, has been demonstrated for the photonic de Broglie wavelength (PBW) over the last two decades [13-15], its potential applications for quantum sensors have still been strictly limited by the N number [15]. Here, a completely different approach to the coherence de Broglie wavelength (CBW) [16,17] based on coherence optics is investigated for the basic physics of quantum entanglement generation in a macroscopic regime, where the macroscopically entangled system is compatible with the N00N state in the PBW. This may contradict our common belief that quantum entanglement cannot be generated by any classical means. Thus, the discussion or argument should be focused on whether coherence is classical or quantum. Compared with classicality defined by individual particles such as coins and dices according to Bell's inequality theorem [18,19], coherence can be either one depending on the choice of phase basis in a coupled system [17]. The present goal is to investigate path superposition in a coupled classical system, where Heisenberg's uncertainty principle is not violated according to Popper's though experiments [20].

Recently, completely different approaches [16,21] for quantum entanglement generation have been suggested to understand secrets of the mysterious quantum phenomena such as anticorrelation in a Hong-Ou-Mandel (HOM) dip [22] and Heisenberg's limit in PBW [13-15]. These new ideas are based on the wave nature of photons, i.e., coherence. The meaning of coherence in this concept is for the first-order intensity correlation in a double-slit or Mach-Zehnder interferometer (MZI) rather than the second-order interference of Hanbury Brown and Twiss (HBT) [23]. Due to coherence, the collective control of an ensemble becomes a basic tool for classical bipartite systems and results in a controllable macroscopic quantum state [24]. Such collective control of an ensemble has already been demonstrated for quantum memories [25-28] and quantum entanglement swapping [29-31]. Unlike these collective systems requiring a preliminarily provided quantum state [24-31], the wave nature-based methods [16,21] only require coherence regardless of a single particle or an ensemble. According to wave-particle duality, the wave nature cannot be compatible with the particle nature simultaneously, even though they can be interchangeable [32]. In other words, the concept of the particle nature must be discarded for the present analysis. Thus, pure coherence optics is applied to a coupled MZI system for the discussion of macroscopic quantum-state generation.

**Results**

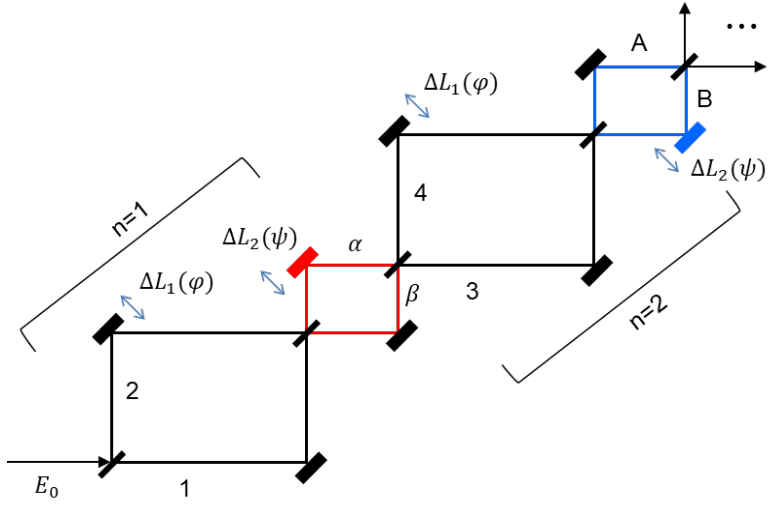

Fig. 1. Schematic of path superposition between two classically independent MZIs. $\varphi = \frac{2\pi}{\lambda}\Delta L_1$; $\psi = \frac{2\pi}{\lambda}\Delta L_2$.

Figure 1 shows a schematic of an n-coupled MZI system for CBW based on phase controls by $\varphi$ and $\psi$, where the basic building block is denoted by n=1. For coupling between independent but identical $\varphi$−MZIs, a $\psi$−MZI is investigated for path superposition between the $\varphi$−MZIs. According to MZI physics, the output fields of the first $\varphi$−MZI whose input field is $E_0$ are distinctively determined by the phase basis of $\varphi \in \{0, \pi\}$:

$$\begin{bmatrix} E_\alpha \\ E_\beta \end{bmatrix} = \frac{1}{2}\begin{bmatrix} e^{i\psi} & 0 \\ 0 & 1 \end{bmatrix}\begin{bmatrix} 1 & i \\ i & 1 \end{bmatrix}\begin{bmatrix} 1 & 0 \\ 0 & e^{i\varphi} \end{bmatrix}\begin{bmatrix} 1 & i \\ i & 1 \end{bmatrix}\begin{bmatrix} E_0 \\ 0 \end{bmatrix}$$
$$= \frac{1}{2}\begin{bmatrix} e^{i\psi}(1-e^{i\varphi}) & ie^{i\psi}(1+e^{i\varphi}) \\ i(1+e^{i\varphi}) & -(1-e^{i\varphi}) \end{bmatrix}\begin{bmatrix} E_0 \\ 0 \end{bmatrix}. \qquad (1)$$

From equation (1), the corresponding intensities represent classical bound governed by the Rayleigh criterion or a diffraction limit, whose phase resolution is $\lambda/2$, where the $\lambda$ is the wavelength of the input field $E_0$ (see the upper panels of Fig. 2):

$$I_\alpha = \frac{1}{2}(1 - cos\varphi)I_0, \qquad (2)$$
$$I_\beta = \frac{1}{2}(1 + cos\varphi)I_0, \qquad (3)$$

where $I_0 = E_0 E_0^*$. Unlike previous analysis with asymmetric coupling between $\varphi$−MZIs with an opposite position of $\Delta L_1$ between n=1 and n=2 MZIs [16], Fig. 1 is set for the same $\Delta L_1$ representing identical MZIs, and an asymmetric $\psi$ is for path superposition between the identical $\varphi$−MZIs. The diffraction limit as a classical bound in an MZI is clear as shown by the phase resolution of $\frac{\lambda}{2}$ (or $\pi$) in the upper right panel in Fig. 2.

For the coupled MZI with the asymmetric $\psi$ as shown in Fig. 1, the output fields are represented as:

$$\begin{bmatrix} E_A \\ E_B \end{bmatrix} = \frac{1}{16}\begin{bmatrix} (1-e^{i\varphi}) & i(1+e^{i\varphi}) \\ ie^{i\psi}(1+e^{i\varphi}) & -e^{i\psi}(1-e^{i\varphi}) \end{bmatrix}\begin{bmatrix} e^{i\psi}(1-e^{i\varphi}) & ie^{i\psi}(1+e^{i\varphi}) \\ i(1+e^{i\varphi}) & -(1-e^{i\varphi}) \end{bmatrix}\begin{bmatrix} E_0 \\ 0 \end{bmatrix}$$
$$= \frac{1}{4}\begin{bmatrix} e^{i\psi}(1-e^{i\varphi})^2 - (1+e^{i\varphi})^2 & i(e^{i\psi}-1)(1-e^{i2\varphi}) \\ i(e^{i\psi}-1)(1-e^{i2\varphi}) & -e^{i\psi}(1-e^{i\varphi})^2 + (1+e^{i\varphi})^2 \end{bmatrix}\begin{bmatrix} E_0 \\ 0 \end{bmatrix}. \qquad (4)$$

The corresponding output intensities of equation (4) are as follows (see the lower panels of Fig. 2):

$$I_A = \frac{1}{2}[1 + cos\varphi^2 + sin\varphi^2 cos\psi], \qquad (5)$$

$$I_B = \frac{1}{4}[(1-cos\varphi)(1-cos\psi)]. \quad (6)$$

From equations (5) and (6), the following $\psi-$basis controls are separately considered to derive the function of phase basis-related superposition (see the lower right panel in Fig. 2):

(i)  For $\psi = 0$,

The output intensities are $\varphi$ independent, resulting in a fixed intensity as shown by the green dotted and solid lines for $I_A = I_0$ and $I_B = 0$, respectively. This identity relation resulting from double unitary transformations has already been applied for a coherence version of quantum key distribution in the name of unconditionally secured classical key distribution [33,34].

(ii)  For $\psi = \pm\pi$,

The output intensities are $\varphi$ dependent, resulting in a swing property between classical and quantum bounds: $I_A = I_0(1 + cos2\varphi)/2$ and $I_B = I_0(1 - cos2\varphi)/2$. In this case, the modulation term $cos2\varphi$ cannot be obtained by classical physics governed by the Rayleigh criterion. With a proper choice of $\psi$ value, the phase resolution of output intensities is doubled due to the $cos2\varphi$ term as shown in the third column of Fig. 2. This nonclassical feature of CBW [16,17] is exactly the same as the PBWs for N=4 (discussed in Fig. 3) [14]. Unlike the particle nature-based PBW, the wave nature-based CBW results from the path superposition between identical MZIs via a dummy MZI. The role of path superposition by $\psi-$MZI in Fig. 1 is unique and unprecedented.

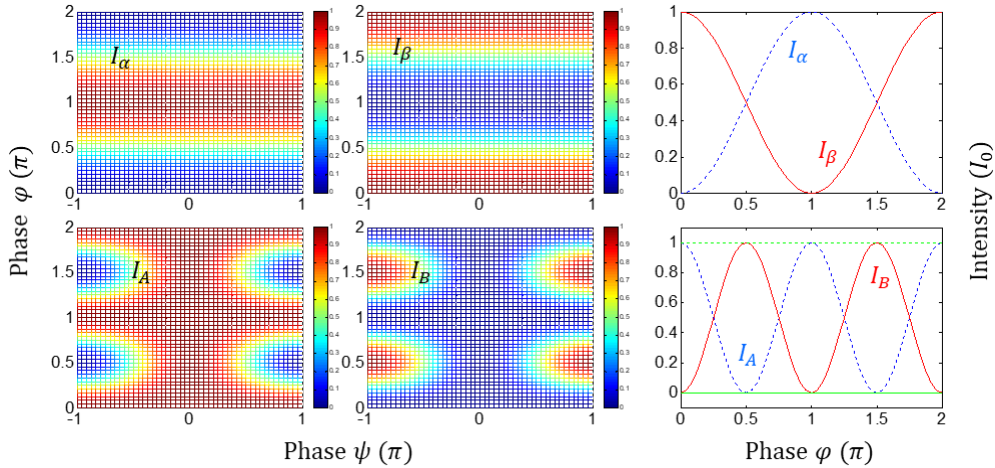

Fig. 2. Numerical calculations for Fig. 1. Upper panels: n=1. Lower panels: n=2. Blue and red curves: $\psi = \pm\pi$. Green dotted (solid) line: $I_A$ ($I_B$) for $\psi = 0$.

(iii)  For $\psi = \pm\pi/2$,

The output intensities are $\varphi$ dependent but do not reach the nonclassical bound, where the phase resolution fluctuates across the diffraction limit (details are analyzed in Fig. 3): $I_A = I_0(1 + cos\varphi^2)/2$ and $I_A = I_0(1 - cos\varphi)/4$. Neither a single MZI nor a coupled MZI without proper path superposition reaches the quantum state at all (see the Supplementary Information).

In a short conclusion, the $2\varphi$ modulation term in case (ii) for $\psi = \pm\pi$ shows a definite quantum feature which cannot be obtained by any classical means. Thus, the function of $\psi-$based superposition between two identical MZIs decides whether the coupled system works for a classical one or a quantum one. The anticorrelation or entanglement condition is satisfied with a proper phase-basis choice. To function as a quantum system, the phase-basis choice must be $\varphi_{AB} \in \{0, \frac{\pi}{2}, \pi, \frac{3\pi}{2}\}$ as shown in the lower right corner of Fig. 2 [17].

Figure 2 shows numerical calculations for equations (4), where the upper panels are for a single MZI (n=1) as a reference, while the lower panels are for a double MZI (n=2) in Fig. 1. For the single MZI, the phase resolution is limited by the Rayleigh criterion at $\lambda/2$. In the double MZI, however, the phase resolution strongly depends on the choice of the phase basis of $\psi$. If $\psi = 0$, the coupled system behaves as a perfect correlation system as shown by the green dotted and solid curves (see the lower panels). If $\psi = \pm\pi$, the coupled system distinctly swings between quantum and classical bounds, representing the nonclassical phase resolution demonstrated in the PBW as shown by the blue dotted and red curves in the lower right corner of Fig. 2. For all other values of $\psi$, the coupled system belongs between the classical and quantum systems.

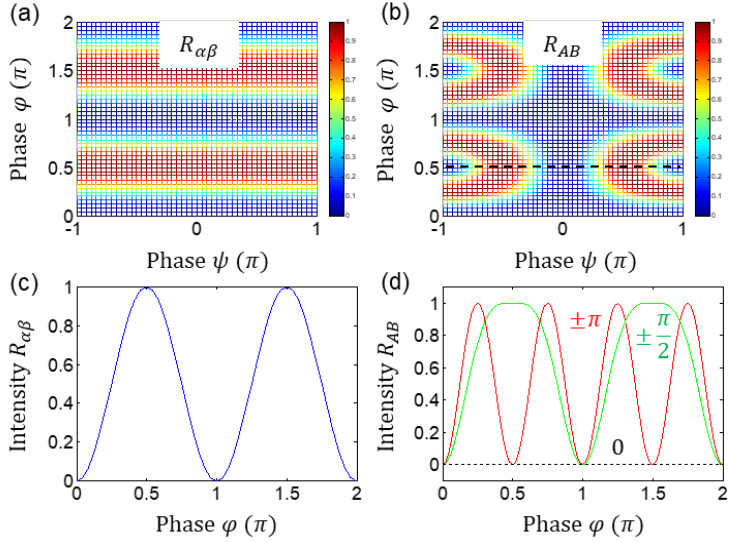

Fig. 3. Numerical calculations for normalized intensity product $R_{ij}$ for Fig. 2. (a) $R_{\alpha\beta}$. (b) $R_{\alpha\beta}$. (c) $R_{\alpha\beta}$ for $\psi = \pm\pi$. (d) $R_{AB}$ for $\psi = \pm\pi$ (red), $\psi = 0$ (dotted), $\psi = \pm\pi/2$ (green). $R_{AB}(\psi)$ for $\varphi = \pm\pi/2$ is the same as (c): dotted line in (b).

Figure 3 shows intensity product $R_{ij}$ ($= 4I_i I_j$) for Fig. 2. Figures 3(a) and (b) show the product $R_{ij}$ as functions of $\varphi$ and $\psi$. Figures 3(c) and (d) are the details of Figs. 3(a) and (b), respectively, where Fig. 3(c) represents the Rayleigh criterion as the conventional coherence limit in a single MZI comparable to N=2 in the PBW [12,14]. The modulation frequency of $R_{AB}$ in Fig. 3(d) varies between the classical and quantum bounds depending on the $\psi$ values as analyzed above in Fig. 2. For $\psi = 0$ (see the dotted line in Fig. 3(d)), $R_{AB} = 0$ shows an extreme bound of perfect correlation, representing a reversible process applied for unconditionally secured classical key distribution [33,34]. In other words, the output direction in the doubly coupled MZI of Fig. 1 is predetermined depending on the $\psi$ basis via double unitary transformations. If $\psi = \pm\pi$, $R_{AB}$ swings between the quantum and classical bounds depending on the $\varphi$ values (see the red curve), resulting in a quantum feature with $\lambda/8$ phase resolution, which is equivalent to the N=4 PBW [14].

If $\psi = \pm\pi/2$, it belongs somewhere between the classical and coherence bounds (see the green curve), partially violating the classical resolution. Unlike the conventional understanding, coherence control in a coupled MZI functions as a decision maker to go for either a classical system or a quantum system. The major discovery in this manuscript is that entanglement represents a maximally correlated system. Thus, the quantum features in a classically coupled MZI system can be manipulated deterministically and macroscopically via path superposition. This macroscopic and deterministic nonclassical property satisfies quantum mechanics in a coupled system [35].

For an extension toward n-coupled MZIs (n=n), the following intensity relations are obtained using equation (4):

$$\begin{bmatrix} E_\alpha \\ E_\beta \end{bmatrix}^{(n)} = \frac{1}{2}(-1)^{n-1}\begin{bmatrix} (1+(-1)^n e^{in\varphi}) & i(1-(-1)^n e^{in\varphi}) \\ -i(1-(-1)^n e^{in\varphi}) & (1+(-1)^n e^{in\varphi}) \end{bmatrix}. \qquad (7)$$

The corresponding intensities of equation (7) are as follows:

$$I_\alpha^{(n)} = \frac{1}{2}I_0[1+(-1)^n \cos(n\varphi)], \qquad (8)$$

$$I_\beta^{(n)} = \frac{1}{2}I_0[1-(-1)^n \cos(n\varphi)]. \qquad (9)$$

In equations (8) and (9), the condition needed for anticorrelation or field bunching is $\varphi = \frac{m\pi}{n}$ (m=0,1,2,…,n), as shown in Figs. 4 and 5. As a result, the principal phase bases of the n-coupled MZI are 0 and $\pi/n$. This relation is equivalent to the enhanced phase resolution in the CBW represented by $\lambda_{CBW} = \frac{\lambda}{2n}$ [16]. As is well understood, the enhanced phase resolution results from the tensor product between n-bipartite MZIs [17]. A proper superposition for the tensor product plays a critical role as shown in Figs. 2-5.

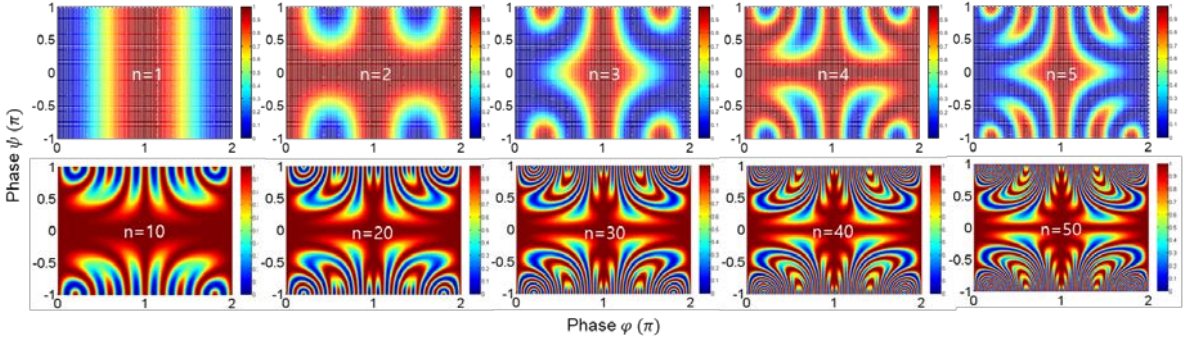

Fig. 4. Numerical calculations for the output field intensity $I_\alpha^{(n)}$ for different n in Fig. 1.

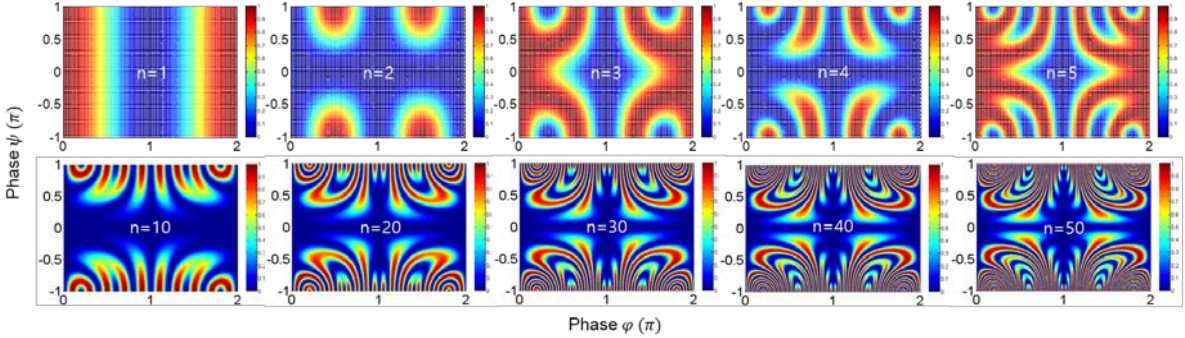

Fig. 5. Numerical calculations for the output field intensity $I_\beta^{(n)}$ for different n in Fig. 1.

The normalized intensity product $R_{\alpha\beta}^{(n)}$ is given by $\langle I_\alpha^{(n)} I_\beta^{(n)} \rangle / \langle I_0 \rangle^2$ (see the Supplementary Information):

$$R_{\alpha\beta}^{(n)} = \frac{1}{2}\langle 1 - cos2n\varphi \rangle. \qquad (10)$$

Unlike the particle nature of photons, the average value in equation (10) is for a single shot measurement. Thus, the phase resolution in the n-coupled asymmetric MZI of Fig. 1 is enhanced by a factor of n, where n=1 is for the Rayleigh criterion or diffraction limit in classical physics (see Fig. 3(c)).

Considering that the intensity correlation of coherent lights is $g^{(2)}(0) = 1$ as determined by Poisson statistics, the normalized intensity product in equation (10) gives different one. A detailed discussion for the normalization method will be given elsewhere. Unlike the general understanding of entangled photon pair-based nonclassical features such as a HOM dip and Bell inequality violation, coherent light can also generate the same quantum features of the PBW as shown in Figs. 4 and 5. These quantum features are due to proper MZI superposition, where the asymmetric $\psi$ plays a key role. The enhanced phase resolution for n ≥ 2 is direct evidence of the quantum features of the CBW based on coherence optics via path superposition. Unlike the PBW, the CBW is deterministic and macroscopic.

In conclusion, the quantum features of the CBW were analyzed in a coupled interferometric scheme of MZIs via superposition control, where the asymmetric coupling method plays an essential role. Due to the coherence optics of MZIs, the generated quantum features were deterministically and macroscopically controlled for on-demand quantum features. Thus, deterministic and macroscopic entanglement generation can be applied for efficient quantum information processing.

**Acknowledgment**

BSH acknowledges that this work was supported by the "Practical Research and Development support program by GTI (GIST Technology Institute)" funded by GIST in 2021.